\documentclass[10pt,prb,aps,twocolumn,floatfix]{revtex4}
\usepackage{graphicx}

\begin{document}



\title{Paramagnonlike excitations and spin diffusion
in magnetic resonance studies \\ of copper oxide superconductors }
\author{Igor A. Larionov\cite{Auth1}}
\address{Magnetic Radiospectroscopy Laboratory, Department of Physics,
Kazan State University, 420008 Kazan, Russia }

Journal Ref.: Physical Review B, accepted, in press

\begin{abstract}

The relaxation function theory for a doped two-dimensional Heisenberg
antiferromagnetic system in the paramagnetic state for all wave vectors
through the Brillouin zone is presented in view of low frequency
response of high-$T_c$ copper oxide superconductors. We deduced the
regions of long lifetime [$T \lesssim$$400(1-4x)$~K] and "overdamped"
[$T \gtrsim$$700(1-4x)$~K] paramagnonlike excitations in the temperature
($T$)-doping index ($x$) phase diagram from plane oxygen nuclear
spin-lattice relaxation rate $^{17}(1/T_1)$ data in up to optimally
doped La$_{2-x}$Sr$_{x}$CuO$_{4}$ thus providing the regimes for the
spin wave concept and the ''overdamped'' mode.

\end{abstract}

\maketitle

\section{INTRODUCTION}

Plane copper oxide high-temperature superconductors (high-$T_c$) are the {\it doped}
$S$=1/2 two dimensional Heisenberg antiferromagnetic (2DHAF) systems.\cite{RMP98} In the
carrier free regime, the elementary excitations are spin waves,\cite{CHN,Takahashi_SW}
magnons in the quasiparticle language, a concept widely known and thoroughly investigated
in the past.\cite{Hone74} Therefore it is tempting to consider the doped 2DHAF systems in
terms of magnonlike excitations (strictly speaking, the paramagnon, a notation used for
spin fluctuations in representation of {\it damped} spin waves) and the magnon lifetime is
characterized by the damping of spin waves. Then the questions arise: what happens with
spin waves when we dope the system? What are the elementary excitations - are they still
paramagnonlike? The motion of charge carriers even in the optimally doped (maximum $T_c$)
high-$T_c$ is known to take place in the presence of strong AF
fluctuations\cite{Moriya_adv_rpp} and spin waves in 2DHAF systems persist even without
long range order in the paramagnetic state,\cite{CHN,Takahashi_SW,Hone74,Tognetti} so the
questions make sense, but, finally, what one can say about the {\it lifetime} of these
excitations when we dope the system?

From experimental point of view the spin-wavelike features have been
revealed\cite{Res_SW} by neutron scattering (NS) even in the nearly
optimally doped YBa$_2$Cu$_3$O$_{6.85}$ and, contrary to the predictions
within the weak coupling theory,\cite{LevErNor} no isotope effect on the
''resonance peak'' (RP) frequency have been
observed\cite{No_O_isotop_res} in YBa$_2$Cu$_3$O$_{6.89}$. The RP
phenomenon disappears in the overdoped phase\cite{Hwang} thus raising
questions about its appearance within the weak coupling
theories\cite{iL05} and AF spin excitations disappear in overdoped
La$_{1.7}$Sr$_{0.3}$CuO$_{4}$ leading to the conclusion that "the AF
spin correlations in superconducting samples must be vestiges of the
parent insulator".\cite{Wakimoto07_x03} Moreover, NS
data\cite{Stock06,Stock07} in underdoped YBa$_2$Cu$_3$O$_{6.35}$ with
$T_c$=18~K shows the {\it commensurate} AF short range order, no
well-defined resonant mode, and similarities between the low-energy
magnetic excitations in YBa$_2$Cu$_3$O$_{6.35}$ and carrier free
insulating 2DHAF, e.g., YBa$_2$Cu$_3$O$_{6.15}$ and
La$_2$CuO$_4$.\cite{Coldea} The spin diffusive contribution
$^{17}(1/T_1)_{\mathit Diff}$ to plane oxygen nuclear spin-lattice
relaxation rate $^{17}(1/T_1)$ cannot be excluded solely by failure to
detect\cite{Thurber_imai_17O_1997} the changes in $^{17}(1/T_1)$ by
varying the nuclear magnetic resonance (NMR) frequency $\omega$ since
$^{17}(1/T_1)_{\mathit Diff}$ varies rather weakly\cite{iL04} with
$\omega$.

In this paper we use the Mori-Zwanzig projection operator
procedure\cite{Mori,Zwanzig} and thus we are unprejudiced regarding the
role of $q \approx 0$ (spin diffusion) and $Q \approx (\pi,\pi)$ wave
vectors in the imaginary part of dynamic spin susceptibility
$\chi''({\bf q},\omega)$ of doped 2DHAF system which is especially
important for $^{17}(1/T_1)$. Spin diffusion is a spatial smoothing of
heterogeneous spin polarization in a system of localized magnetic
moments and in the presence of strong damping (short magnon lifetime)
the spin dynamics changes from wavelike to diffusive. The approximations
we use for the relaxation function are within the Markovian
approximation and ''by itself the Markovian situation can be valid even
in the absence of any picture of the system in terms of well-defined
excitations''.\cite{Tognetti} We will emphasize the spin-wavelike
features in $\chi''({\bf q},\omega)$ of copper oxide high-$T_c$ and
extract the lifetime of spin-wavelike excitations from $^{17}(1/T_1)$
data.

\section{BASIC RELATIONS}

We start from the $t$-$J$ Hamiltonian \cite{Anderson_t_J} known as the
minimal model for the electronic properties of high-$T_c$ cuprates
\begin{equation} \label{Ht_J}
H_{t-J}= \sum_{i,j,\sigma}t_{ij}X_i^{\sigma 0}X_j^{0\sigma} + J\sum_{i>j}
({\bf S}_i{\bf S}_j-\frac{1}{4}n_i n_j ) ,
\end{equation}
written in terms of the Hubbard operators $X_i^{\sigma 0} $ that create
an electron with spin $\sigma$ at site $i$ and ${\bf S}_i$ are spin-1/2
operators. Here, the hopping integral $t_{ij}$=$t$ between the nearest
neighbors (NN) describes the motion of electrons causing a change in
their spins and $J$=0.12~eV is the NN AF coupling constant. The spin and
density operators are defined as follows: $S_i^{\sigma}$=$X_i^{\sigma
\tilde \sigma}$, $S_i^z $=(1/2)$\sum_\sigma \sigma X_i^{\sigma \sigma}$,
$n_i$=$\sum_\sigma X_i^{\sigma \sigma}$, ($\sigma $=$-\tilde
\sigma$), with the standard normalization $X_i^{00}+X_i^{++}+X_i^{--}$=1.

We formulate our study of the spin fluctuations following
Mori,\cite{Mori} who showed it's efficiency for both the classical (and
essential equivalence to Brownian motion) and quantum (e.g., Heisenberg
systems of arbitrary dimension) many body systems.\cite{Tognetti} The
time evolution of a dynamical variable $ S^z_{\bf k} (\tau )$, say, is
given by the equation of motion,
\begin{equation} \label{Liouville}
\dot{S}^z_{\bf k}(\tau ) \equiv \frac{d S^z_{\bf k}(\tau )}{d \tau} =
i{\mathcal{L}} S^z_{\bf k}(\tau ) \rightarrow [H_{t-J}, S^z_{\bf k}
(\tau )],
\end{equation}
where the Liouville operator ${\mathcal{L}} $ in the quantal case
represents the commutator with the Hamiltonian. The projection of the
vector $ S^z_{\bf k} (\tau )$ onto the $S^z_{\bf k} \equiv S^z_{\bf
k}(\tau$=0) axis, ${\mathcal{P}}_0 S^z_{\bf k}(\tau )={\mathcal{R}}({\bf
k},\tau ) S^z_{\bf k}$, defines the linear projection Hermitian operator
${\mathcal{P}}_0$. One may separate $S^z_{\bf k} (\tau )$ into the
projective and vertical components $ S^z_{\bf k}(\tau ) =
{\mathcal{R}}({\bf k},\tau ) S^z_{\bf k} + (1-{\mathcal{P }}_0)S^z_{\bf
k}(\tau ) $ with respect to the $S^z_{\bf k}$ axis, where
${\mathcal{R}}({\bf k},\tau ) \equiv (S^z_{\bf k}(\tau ),(S^z_{\bf
-k})^{*}) (S^z_{\bf k},(S^z_{\bf -k})^{*})^{-1} $ is the relaxation function in the
inner-product notation: $(S^z_{\bf k}(\tau ),(S^z_{\bf -k})^{*}) \equiv
k_B T
\int_0^{1/k_B T} d \varrho \langle e^{\varrho H} S^z_{\bf k}(\tau )e^{-\varrho
H}(S^z_{\bf -k})^{*} \rangle$, and the angular brackets denote the
thermal average.

One may construct a continued fraction representation for the Laplace
transform of the relaxation function, for which Lovesey and Meserve
\cite{LoveseyMeserve,Tognetti} used a three pole approximation, $
{\mathcal{R}}^L({\bf k},s)$=$\int_0^{\infty} d
\tau \mbox{\hspace{1mm}} e^{-s \tau} {\mathcal{R}} ({\bf k},\tau )
$$\approx$$ 1/ \{s + \Delta^2_{1{\bf k}}/[s+\Delta^2_{2{\bf k}}/(s+1/
\tau_{\bf k})] \}$, with a cutoff characteristic time $\tau_{\bf k}=
\sqrt { 2 / (\pi \Delta^2_{2{\bf k}} )
} $, by arguing that $S^z_{\bf k}(\tau )$ fluctuations are weakly
affected by higher order random forces. For the relaxation shape
function ${\mathcal{F}}( {\bf k }, \omega )$=$ Re [ {\mathcal{R}}^L
({\bf k},i \omega)]/\pi$, this gives
\begin{equation} \label{FkwD1D2}
{\mathcal{F}}({\bf k},\omega)=\frac{\tau_{\bf k} \Delta^2_{1{\bf k}}
\Delta^2_{2{\bf k}}/\pi} {[\omega \tau_{\bf k} (\omega^2-\Delta^2_{1{\bf
k}}-\Delta^2_{2{\bf k}})]^2 + (\omega^2 - \Delta^2_{1{\bf k}})^2},
\end{equation}
where $\Delta_{1 {\bf k}}^2$ and $\Delta_{2 {\bf k}}^2$ are related to
the frequency moments, $\left< \omega^n_{\bf k} \right> =
\int_{-\infty}^{\infty}
 d \omega \mbox{\hspace{1mm}} \omega^n {\mathcal{F}}({\bf k},\omega) =
(1/i^n) \left[ d^n {\mathcal{R}}({\bf k},\tau ) / d \tau^n \right]
_{\tau =0} $, of ${\mathcal{R}}({\bf k},\tau )$ as $\Delta^2_{1{\bf
k}}=\left< {\omega^2_{\bf k}} \right>$, $\Delta^2_{2{\bf k}}=(\left<
{\omega^4_{\bf k}} \right> /\left< {\omega^2_{\bf k}} \right> )-\left<
{\omega^2_{\bf k}} \right> $ for $\tau \gtrsim \tau_{\bf k}$. Note that
${\mathcal{F}}({\bf k},\omega)$ is normalized to unity
$\int_{-\infty}^{\infty} d \omega {\mathcal{F}}({\bf k},\omega)=1$ and
is even in both ${\bf k}$ and $\omega$. The expression for the second
moment, $\left< {\omega^2_{\bf k}} \right> = i \langle [\dot{S}^z_{\bf
k}, S^z_{-{\bf k}} ] \rangle / \chi({\bf k})=-(8J c_1 - 4 t_{\mathit
eff} T_1 ) (1-\gamma_{\bf k} ) / \chi({\bf k})$, is compact, while
$\left< {\omega^4_{\bf k}} \right> = i\langle[ \ddot{S}^z_{\bf k},
\dot{S}^z_{-{\bf k}} ] \rangle  / \chi({\bf k}) $ is cumbersome and is
not reproduced here (see Ref.~\onlinecite{iL04} for details).

The static spin susceptibility has been {\it derived} within the $t$-$J$
model in the overall temperature and doping range\cite{Zav98}
\begin{equation} \label{StatSusc}
\chi ({\bf k}) = \frac {4 | c_1 |}{J g_{-} (g_{+} + \gamma_{\bf k})},
\end{equation}
and has the same structure as in the isotropic spin-wave
theory.\cite{Sokol_spin_wave} The parameter $g_{+}$ is related to
correlation length $\xi $ via the expression $\xi
/a $=$1/(2 \sqrt{ g_{+}-1}$), where $a$=3.8~\AA is a lattice unit.
The transfer amplitude between the NN is given by: $T_1 \equiv - (1/4)
\sum_{\rho} \left< X_{i}^{\sigma 0} X_{i+\rho}^{0 \sigma} \right>
$=$ p\sum_{\bf k} \gamma_{\bf k} f_{\bf k}^h$, the index $\rho$ runs
over NN, $ \gamma_{\bf k} $=$ (1/4) \sum_{\bf \rho}
\exp(i{{\bf k} \rho}) $=$ (1/2) (\cos k_x a +\cos k_y a)$, and
$f_{\bf k}^h $=$[\exp(-E_{\bf k}+\mu )/k_BT+1]^{-1}$ is the Fermi
function of holes. The number of {\em extra} holes, due to doping,
$\delta$, per one plane Cu$^{2+}$, can be identified with the Sr content
$x$ in La$_{2-x}$Sr$_{x}$CuO$_{4}$. The chemical potential $\mu$ is
related to $\delta$ by $\delta$=$p \sum_{\bf k} f_{\bf k}^h $, with
$p$=(1+$\delta$)/2. The excitation spectrum of holes is given by $E_{\bf
k} $=$ 4 t_{\mathit eff} \gamma_{\bf k}$, where the hoppings, $t$, are
affected by electronic and AF spin-spin correlations $c_1$, resulting in
{\it effective} values,\cite{Anderson_t_J,MyZavDB,Plakida} for which we
set $t_{\mathit eff} = \delta J$/0.2 in order to match the
insulator-metal transition.

For low temperature behavior we use the expression, resulting in {\it
effective} correlation length $\xi_{\mathit eff} $, given
by\cite{iL04,MyZavDB,Keimer1992}
\begin{equation}\label{xi_eff}
\xi_{\mathit eff}^{-1}=\xi_0^{-1}+\xi^{-1}.
\end{equation}
Here, $\xi$ is affected by doped holes, in contrast with the Keimer {\it
et al.}\cite{Keimer1992} empirical equation, where $\xi $ is given by
the Hasenfratz-Niedermayer formula and there was no influence of the
hole subsystem on $\xi$. Thus from now on we replace $\xi$ by
$\xi_{\mathit eff}$. For doped systems we use the explicit
expression\cite{Zav98} for $\xi$ which is much more complicated compared
with simple relation $ \xi /a \simeq  (J \sqrt{\mbox{\hspace{1mm}} g_{-}
} /k_B T) \exp (2\pi
\rho_S/k_B T)$, valid for carrier free or lightly doped systems.\cite{iL04,Zav98} In the
best fit of $\xi_{\mathit eff}$ to experimental
data\cite{Keimer1992,Ae97} the relation $\xi_0 = a/n_{\xi} x$ is most
suited\cite{iL05,iL04} which one may attribute to stripe picture, where
$n_{\xi}$=2 for $x \leq$0.05 and $n_{\xi} $=1 near optimal
($x\approx$0.15) doping. The results of the calculations are summarized
in Table~I. We consider here the case of La$_{2-x}$Sr$_{x}$CuO$_{4}$
with the simplest crystalline structure for brevity and luck to thorough
experimental data set.

\section{Plane oxygen nuclear spin-lattice relaxation}

The plane oxygen nuclear spin-lattice relaxation rate $^{17}(1/T_1 )$
has three contributions:
\begin{equation} \label{T1_x_sd_L}
^{17}(1/T_1 )={^{17}}(1/T_1 )_{sw} + {^{17}}(1/T_1)_{Korr} + {^{17}}(1/T_1)_{\mathit Diff}.
\end{equation}
The contribution from spin-wavelike excitations is given by
\begin{equation} \label{T1ABSkw}
^{17}\left( 1/T_1 \right)_{sw} =\frac{2 k_B T}{\omega_{0} } \sum_{|{\bf k}| > 1/\xi_{\mathit eff}}
{ ^{17} F ({\bf k})^2 }  \chi_{L}'' ({\bf k},\omega_{0}),
\end{equation}
where $\omega_0 = 2\pi \times$52~MHz $\simeq$2.15$ \times 10^{-4}$~meV
($\ll T, J$) is the measuring NMR frequency at 9~Tesla. The quantization
axis is along the crystal $c$ axis and the wave vector dependent
hyperfine form factor for plane $^{17}$O sites is given by $^{17} F
({\bf k})^2 = 2 C^2 (1+ \gamma_{\bf k} )$, with $C =2.8 \times
10^{-7}$~eV.\cite{Pines_NAFL}

\begin{table}

{ TABLE~I.} The calculated in the $T \rightarrow$0 limit NN AF spin-spin correlation
function $c_1$=(1/4)$\sum_\rho \langle S_i^zS_{i+\rho}^z\rangle$, the parameter
$g_{-}$, the spin stiffness constant $\rho_S$ using the expressions and the procedure
as described in Refs.~\onlinecite{Zav98} and \onlinecite{MyZavDB}, the calculated spin
diffusion constant, $D$, following Ref.~\onlinecite{iL04} together with the values of
Korringa-type contribution constant $K_{K}$ and the spin-wavelike damping
renormalization constant ${\Gamma_{r}}$ as extracted from comparison with
$^{17}(1/T_1)$ NMR data.

$\phantom{\frac{\mbox{\hspace{1mm}}}{\mbox{\hspace{1mm}}}}$ {
\begin{tabular}
{l c c c c c c c c} {$x$} & $c_1$ & ${ g_{-} }$ & $2\pi \rho_S / J$ & ${\mathit
D}/Ja^2$ &  $K_{K}$, (sK)$^{-1}$ & $ {\Gamma_{r}} $, K$^{-3}$
\\
\hline

 0.025 & $-$0.1133 & 4.102 & 0.36  &  2.60 & 0.023 &  4.1$\times$10$^{-9}$
\\

 0.035 & $-$0.1115 & 4.060 & 0.35  &  2.54 & 0.024 & 5.5$\times$10$^{-9}$
\\

 0.05 & $-$0.1018 & 3.827 & 0.285  &  2.47 & 0.051 & 7.5$\times$10$^{-9}$
\\

 0.115 & $-$0.0758 & 3.252 & 0.2  &  3.51 & 0.147 & 32$\times$10$^{-9}$
\\

 0.15 & $-$0.0617 & 2.947 & 0.13  &  3.81 & 0.215 & 41$\times$10$^{-9}$

\end{tabular}
}
\end{table}

The contribution from itinerant holes, of Korringa type, $^{17}(1/T_{1}
)_{Korr} = K_{K} T$, should grow with doping $x$ and will be the
adjustable parameter. The contribution from spin diffusion (small wave
vectors ${\bf k} < 1/\xi_{\mathit eff}$) may be calculated from general
physical grounds, namely, the linear response theory, hydrodynamics, and
fluctuation-dissipation theorem\cite{ForsterBook} (see also
Ref.~\onlinecite{iL04})
\begin{equation} \label{T1_Diff}
^{17} ( 1/T_1 )_{\mathit Diff}=\frac {^{17} F(0)^2 k_B T a^2 \chi({\bf k}=0)} {\pi \hbar D} {\Lambda },
\end{equation}
where ${\large \Lambda }$=$[1/(4\pi)]\ln [1+D^2/(\omega_0^2
\xi_{\mathit eff}^4)]$ and the calculated values of spin diffusion
constant, $D$, are given in Table~I.

Since the relaxation function can be understood within the spin-wave
framework,\cite{Tognetti} the temperature and doping dependence of the
{\it damping } of the spin-wavelike excitations may be studied further.
The spin-wavelike dispersion, renormalized by interactions, is given by
the relaxation function,\cite{Tognetti}
\begin{equation}\label{w_SW_Fkw}
\omega^{sw}_{\bf k} = 2\int_0^{\infty} d \omega
\mbox{\hspace{1mm}} \omega {\mathcal{F}}({\bf k},\omega),
\end{equation}
where the integration over $\omega$ in Eq.~(\ref{w_SW_Fkw}) has been
performed analytically and exactly. We assume the Lorentzian form of the
imaginary part of the dynamic spin susceptibility,
\begin{equation} \label{chi_def}
\chi_L '' \left( {{\rm {\bf k}},\omega } \right) = {\frac{{\chi} \left( {\rm {\bf k}}
\right) \omega \Gamma_{\rm {\bf k}} }{[\omega - \omega_{\rm {\bf k}}^{sw} ]^2 +
\Gamma_{\rm {\bf k}}^2 } + \frac{{\chi } \left( {\rm {\bf k}} \right) \omega
\Gamma_{\rm {\bf k}}}{[\omega +\omega_{\rm {\bf k}}^{sw} ]^2 +\Gamma_{\rm {\bf k}}^2}},
\end{equation}
for ${\bf k}$ around $(\pi,\pi)$. We accept to the leading order the
cubic temperature dependence\cite{Kopietz,Tyc} for the damping of
spin-wavelike excitations $\Gamma_{\rm {\bf k}} $=$
\Gamma_{r} T^3 \eta_{\rm {\bf k}} $, where the wave vector dependence is
given by $\eta_{\rm {\bf k}} $=$ \sqrt {\langle \omega_{\rm {\bf k}}^2
\rangle - (\omega_{\rm {\bf k}}^{sw})^2 } $.

\section{RESULTS AND DISCUSSION}

\begin{figure} [tbp]
 \centering \includegraphics[width=1.15\linewidth] {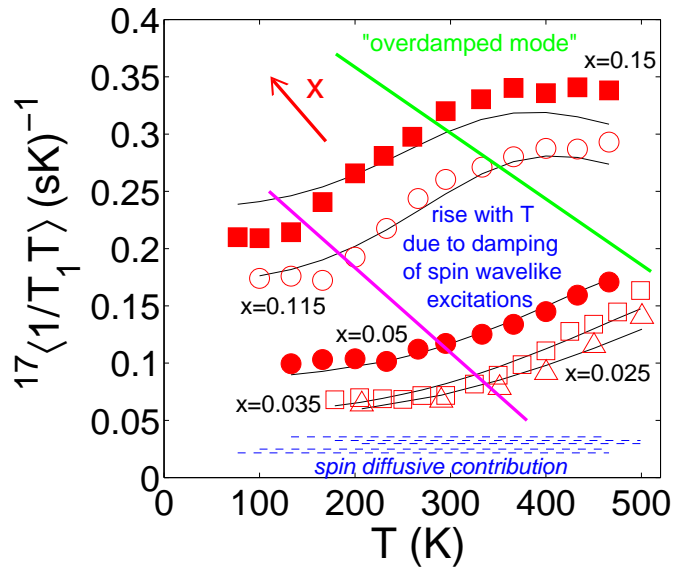}
\caption{(Color online) Temperature and doping behavior of plane oxygen $^{17}(1/T_1 T)$
for La$_{2-x}$Sr$_{x}$CuO$_{4}$. Dashed lines show the spin diffusive
contribution. Thin solid curves are the fits by Eq.~(\ref{T1_x_sd_L}) to
NMR data\protect\cite{Thurber_imai_17O_1997,Singer_imai_17O_2005} as
described in the text. Thick solid lines separate schematically the
regions of lightly damped (low T region), damped, and overdamped
magnon-like excitations. }
\label{rel_17O}
\end{figure}

Figure~\ref{rel_17O} shows the plane oxygen $^{17}( 1/T_1 )$ fitted by
Eq.~(\ref{T1_x_sd_L}) with two adjustable parameters: $K_K$ and
$\Gamma_r$, which values are given in Table~I. The quality of the fit is
very good, which we treat as the validity of our theory. The importance
of $^{17}(1/T_{1} )_{Korr}$ and $^{17}(1/T_{1} )_{\mathit Diff}$ in
plane oxygen $^{17}( 1/T_1 )$, in contrast\cite{iL04,MyZavDB} with plane
copper $^{63}( 1/T_1 )$, is due to the filtering of the $(\pi,\pi)$
contribution by the plane oxygen hyperfine form factor. Obviously, in
the absence of $^{17}(1/T_{1} )_{Korr}$ and $^{17}(1/T_{1} )_{\mathit
Diff}$, it is hard to explain the measured $^{17}(1/T_{1} )$ at large
$x$ with any form of the damping function. In general, the damping grows
with doping $x$, as it should. It should be emphasized that the increase
of $^{17}( 1/T_1 )$ with temperature is caused by the increase of the
damping, $\Gamma_{\rm {\bf k}}$. The low $T$ region of lightly damped
paramagnonlike excitations, where the data may be explained by the
theory that neglects the damping,\cite{iL04} is quantified through the
relation $\Gamma_{r} T^3 <$0.2.

\begin{figure}  [tbp]
 \centering \includegraphics[width=1.19\linewidth] {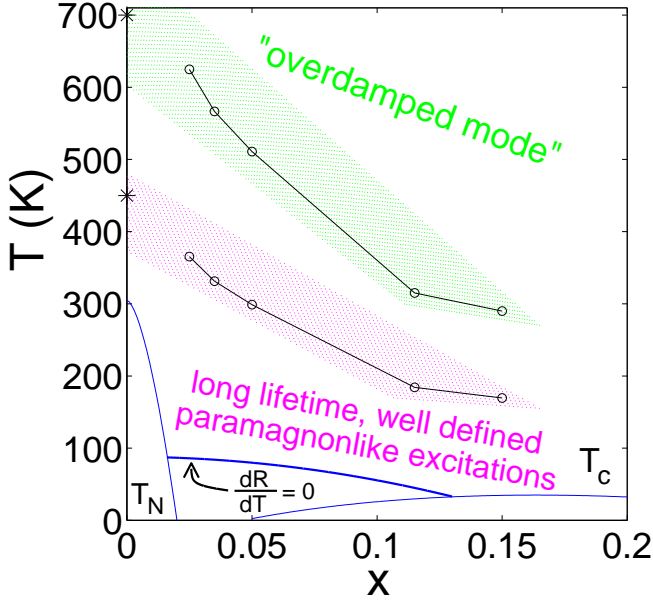}
\caption{(Color online) La$_{2-x}$Sr$_{x}$CuO$_{4}$ phase diagram.
Solid curves show the Neel temperature ($T_N$), the superconducting dome
($T_c$), and dR/dT=0 curve indicates the gradual crossover from
insulating to metallic behaviour from resistivity
measurements.\protect\cite{RMP98} Lower and upper shaded lines with
circles have been extracted from $^{17}(1/T_1 T)$ data shown in
Fig.~\ref{rel_17O} with the conditions ${\Gamma_{r}}T^3$=0.2 and
${\Gamma_{r}}T^3$=1, where $\Gamma_r$ is given in Table~I, may be
approximated by $T\approx$400(1$-$4$x$) and $T\approx$700(1$-$4$x$),
respectively, and separate the regions of lightly damped (long lifetime,
well defined), damped, and overdamped paramagnonlike excitations. The
asterisks mark approximately the corresponding temperatures as extracted
from the data analysis\protect\cite{Thurber_imai_17O_1997} for
Sr$_2$CuO$_2$Cl$_2$.}
\label{ph_diag}
\end{figure}

Figure \ref{ph_diag} shows the temperature ($T$)-doping index ($x$)
phase diagram with the spin-wavelike damping regimes deduced from plane
oxygen nuclear spin-lattice relaxation rate $^{17}(1/T_1)$ data. It is
tempting to speculate that the doping and temperature behavior of these
curves resembles the characteristic "pseudogap" temperatures.

 The form of $\chi_L '' \left( {{\rm {\bf k}},\omega }
\right)$ in Eq.~(\ref{chi_def}) gives the {\it commensurate } response
at low $\omega$ and the incommensurate response at high $\omega$ in
agreement with NS studies in the lightly doped
regime.\cite{RMP98,Stock06,Stock07} Very recently, Stock {\it et
al.}\cite{Stock06,Stock07} reported the evidence for spin waves from NS
studies of underdoped YBa$_2$Cu$_3$O$_{6.35}$ with $T_c$=18~K, where the
magnetic excitations are very similar to that of carrier free 2DHAF
systems. These observations, together with the undoubtful evidence for
disappearance of AF spin excitations in the overdoped
regime,\cite{Wakimoto07_x03} show that AF spin excitations in the
overall doping range of copper oxide high-$T_c$ emanate from those of
the parent insulator, i.e., the spin waves.

\begin{figure}  [tbp]
\centering \includegraphics[width=1.16\linewidth] {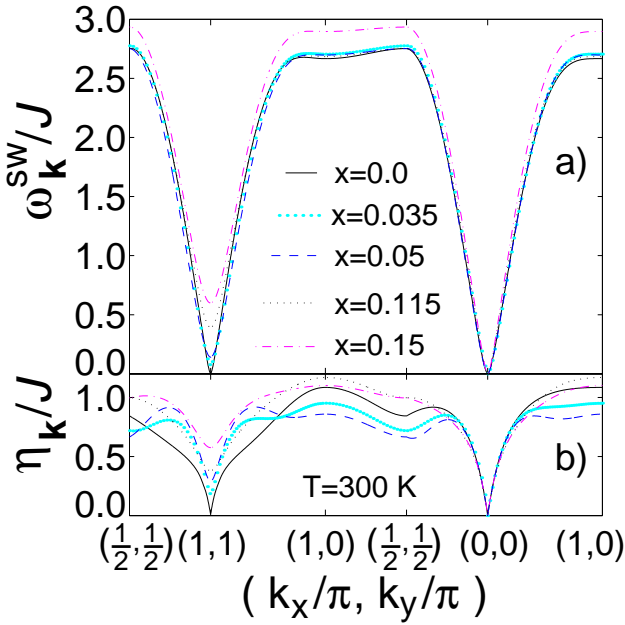}
\caption{(Color online) (a) Spin-wavelike dispersion $\omega^{sw}_{\bf k}$
 and (b) damping function $\eta_{\rm {\bf k}}$ at various dopings along
($\frac 12$,$\frac 12$)-(1,1)-(1,0)-($\frac 12$,$\frac 12$)-(0,0)-(1,0)
route in the Brillouin zone.}
\label{SW}
\end{figure}

The wavevector dependence of spin-wavelike dispersion $\omega^{sw}_{\bf
k}$ and damping $\eta_{\rm {\bf k}}$ is shown in Figs.~\ref{SW}(a) and
\ref{SW}(b), respectively, for various doping levels. Both
$\omega^{sw}_{\bf k}$ and $\eta_{\rm {\bf k}}$ show negligible
temperature dependence below $T$$<$$J/2$ except the region around
($\pi$,$\pi$). For $x$=0, our calculated magnon energy,
$\omega^{sw}_{\bf k}$, at ($\pi$,0) is 3\% lower than at
($\pi$/2,$\pi$/2) in qualitative agreement with Monte Carlo simulations
and series expansion calculations,\cite{Singh} and is similar to that in
Sr$_2$Cu$_3$O$_4$Cl$_2$,\cite{Kim_SrCl_SW} however, is a bit different
from NS data in La$_2$CuO$_4$.\cite{Coldea} The dispersion of
$\omega^{sw}_{\bf k}$ remains approximately the same and the wave vector
dependence of the damping function, $\eta_{\rm {\bf k}}$, in contrast,
possesses significant changes with doping. We emphasize that below
$T\approx$400(1$-$4$x$)~K, where $\Gamma_{r} T^3<$0.2, the damping
$\Gamma_{\rm {\bf k}}$ is much smaller compared with $\omega^{sw}_{\bf
k}$, thus the spin-wavelike excitations are indeed well defined (long
lifetime). Above $T\approx$700(1$-$4$x$)~K, where $\Gamma_{r} T^3 >$1,
the damping, $\Gamma_{\rm {\bf k}}$, becomes compatible with
$\omega^{sw}_{\bf k}$, and grows further with $T$ and $x$ thus providing
the ''overdamped'' mode region.\cite{Prelovsek,MorrPinesSF}

It should be mentioned that the so-called anomalous ''1/8'' doping
problem can be viewed as a consequence of the particular parameter set
at $x\simeq 1/8$ within an extended $t$-$J$ model with the Coulomb
repulsion between the NN and taking into account the polarization of NN
copper spins around copper-oxygen singlet that gives the in-phase domain
structure for the ''stripe'' picture.\cite{iL_MV_1_8} The in-phase
domain is favorable, also in view of the experimental
data,\cite{Teit,Lavrov} compared with the anti-phase domain
model.\cite{Tranquada} This particular case is beyond our present
consideration because of its narrowness.

\begin{figure}  [tbp]
 \centering \includegraphics[width=0.950\linewidth] {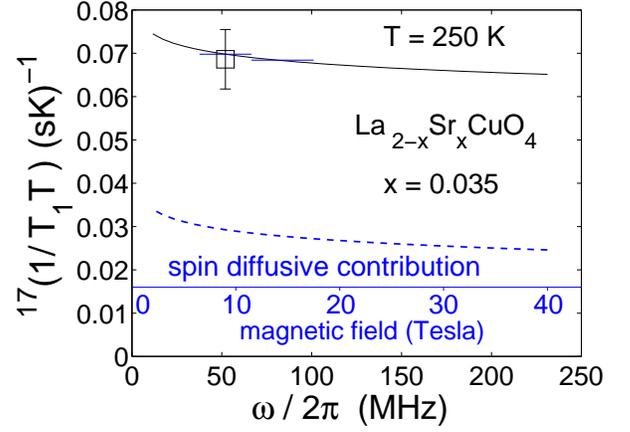}
\caption{(Color online) The calculated field (frequency) dependence of
$^{17}(1/T_1 T)$ (solid curve). The dashed curve shows the spin diffusive
contribution. The error bar shows the size of the symbol (error bar) in
Fig.~\ref{rel_17O}. Short solid horizontal lines indicate the calculated values of
$^{17}(1/T_1 T)$ at the fields 9~T and 14.1~T. }
\label{rel_17O_diff}
\end{figure}

Figure~\ref{rel_17O_diff} and Eq.~\ref{T1_Diff} show that in doped 2DHAF
system the spin diffusive contribution $^{17}(1/T_{1} )_{\mathit Diff}$
scales with the NMR frequency $\omega$ as $\ln(const J/\omega)$, which
is very weak in view of the extraordinary large superexchange coupling
constant $J$. We argue that at low doping level ($x$$\approx$0.03 per Cu
site in La$_{2-x}$Sr$_{x}$CuO$_{4}$) and temperature $T$$<$300~K,
$^{17}(1/T_1)$ is strongly affected by $^{17}(1/T_{1} )_{\mathit Diff}$
luck of the filtering of ($\pi,\pi$) contribution by the oxygen
formfactor. Only a new accurate NMR experiments at very low and very
high magnetic fields may uncover $^{17}(1/T_{1} )_{\mathit Diff}$
contribution to $^{17}(1/T_1)$.

\section{Conclusion}

We applied the relaxation function theory for doped 2DHAF system in the
paramagnetic state and deduced the lifetime of spin-wavelike
excitations, its evolution with doping, from plane oxygen $^{17}(1/T_1)$
NMR data in the underdoped high-$T_c$ layered copper oxides. It is shown
that the spin-wavelike theory is able to reproduce the main features of
low frequency spin dynamics in the normal state of high-$T_c$ cuprates
as observed experimentally. We identified the regions of long lifetime
[$T\lesssim$400(1$-$4$x$)~K] and "overdamped"
[$T\gtrsim$700(1$-$4$x$)~K] paramagnonlike excitations in the
temperature ($T$)-doping index ($x$) phase diagram in up to optimally
doped La$_{2-x}$Sr$_{x}$CuO$_{4}$. The results indicate that
spin-wavelike excitations are indeed a good description of the
quasiparticle excitations even for {\it strongly doped} high-$T_c$
layered cuprates at low temperatures, $T\lesssim$400(1$-$4$x$)~K.

\section{Acknowledgments}

It is a pleasure to thank Peter Fulde and numerous colleagues for
discussions and hospitality at MPI-PKS, Dresden, Germany, and Takashi
Imai for providing with $^{17}(1/T_1)$ NMR data in the electronic form.

\end{document}